\documentclass[letterpaper]{article} 
\usepackage[draft]{aaai25}  
\usepackage{times}  
\usepackage{helvet}  
\usepackage{courier}  
\usepackage[hyphens]{url}  
\usepackage{graphicx} 
\usepackage{natbib}  
\usepackage{caption} 

\usepackage{algorithm}
\usepackage{algorithmic}
\usepackage{amsmath}
\usepackage{newfloat}
\usepackage{listings}
\usepackage[T1]{fontenc}
\usepackage[utf8]{inputenc} 
\DeclareCaptionStyle{ruled}{labelfont=normalfont,labelsep=colon,strut=off} 
\floatstyle{ruled}
\newfloat{listing}{tb}{lst}{}
\floatname{listing}{Listing}
\title{Frechet Music Distance: A Metric For Generative Symbolic Music Evaluation}
\author {
    Jan Retkowski, 
    Jakub Stępniak, 
    Mateusz Modrzejewski
}
\affiliations {
    Institute of Computer Science, Warsaw University of Technology\\
    jan.retkowski.stud@pw.edu.pl,
    jakub.stepniak2.stud@pw.edu.pl, mateusz.modrzejewski@pw.edu.pl
}

\begin{document}

\maketitle

\begin{abstract}
In this paper we introduce the Frechet Music Distance (FMD), a novel evaluation metric for generative symbolic music models, inspired by the Frechet Inception Distance (FID) in computer vision and Frechet Audio Distance (FAD) in generative audio. FMD calculates the distance between distributions of reference and generated symbolic music embeddings, capturing abstract musical features. We validate FMD across several datasets and models. Results indicate that FMD effectively differentiates model quality, providing a domain-specific metric for evaluating symbolic music generation, and establishing a reproducible standard for future research in symbolic music modeling.
\end{abstract}

%
\begin{links}
    \link{Code}{https://github.com/jryban/frechet-music-distance}
\end{links}

\section{Introduction}

Generative symbolic music has emerged as a significant area of research within artificial intelligence, aiming to autonomously create structured, coherent, and expressive compositions in symbolic formats such as MIDI. Unlike audio-based music generation, which directly produces sound waveforms, symbolic music generation operates at an abstract, event-based level, encoding musical elements like pitch, duration, dynamics, and timing. 
The advancements of neural architectures, particularly recurrent neural networks, variational autoencoders and transformers, has accelerated the development of models capable of learning and generating complex symbolic sequences that exhibit stylistic diversity and musicality \cite{sturm2016music} \cite{huang2018music} \cite{shih2022theme}. However, evaluating the quality of generated symbolic music remains challenging due to the multifaceted nature of musical perception and the lack of objective, domain-specific metrics \cite{ji2023survey}. 

Previous evaluation approaches often rely on subjective judgments or low-level statistical metrics, which can be labor-intensive and may not capture the full depth of musical coherence and diversity. To address this gap, we propose the Frechet Music Distance (FMD), a metric specifically designed for symbolic music, providing a scalable and reproducible standard for evaluating generative models in this field.

Our contributions may be summarized as follows:
\begin{itemize}
    \item we introduce FMD, a new metric for evaluating generative symbolic music;
    \item we evaluate FMD with several experiments and show its potential for capturing musical characteristics and usefulness in qualitative measurments of symbolic music that were previously not available;
    \item we release a Python toolkit for computing FMD.
\end{itemize}

To the best of our knowledge, FMD is the first metric designed specifically to evaluate symbolic music generation by comparing distributions of embeddings aimed at capturing essential musical features. Unlike existing metrics that may rely on surface-level statistical comparisons or require extensive human evaluation, FMD leverages learned embedding spaces to provide an objective, reproducible assessment of musical quality and diversity, tailored to the requirements of symbolic music. 


\section{Related work}

\subsection{Frechet Distance metrics}

\begin{figure*}[ht]
\centering
\includegraphics[width=0.8\textwidth]{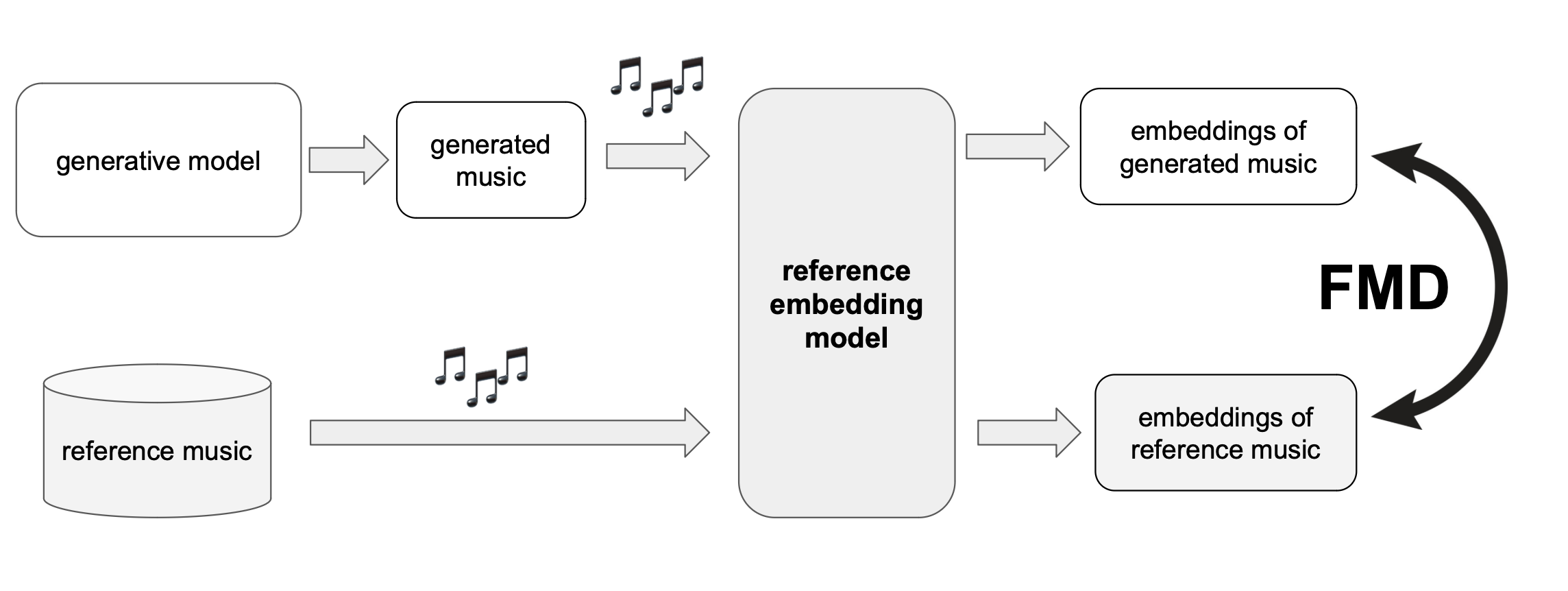} 
\caption{Schematic overview of Frechet Music Distance computation between multivariate Gaussians estimated on the embeddings of generated symbolic music and embeddings of a reference set of music.}
\label{fmd_schema}
\end{figure*}

Frechet distances \cite{dowson1982frechet} are computed between multivariate Gaussians. In the context of machine learning evaluation, the distributions \(\mathcal{N}(\mu_r, \Sigma_r)\) and \(\mathcal{N}(\mu_t, \Sigma_t)\) are estimated from two sets of embeddings: the reference set and the test set. The reference set typically represents a ground truth or real-world data distribution, while the test set is often generated by a generative model. These embeddings are derived using a predefined reference representation model, as illustrated in Figure~\ref{fmd_schema}. The Frechet distance is then calculated as:

\[
\mathrm{FD} = ||\mu_r - \mu_t||^2 + \text{Tr}\left(\Sigma_r + \Sigma_t - 2\sqrt{\Sigma_r \Sigma_t}\right)
\]

where $Tr$ is the matrix trace.

Evaluating generative models presents unique challenges due to the need to measure not only statistical accuracy but also abstract qualities like coherence and perceptual realism. The Frechet Inception Distance (FID) was proposed as an objective metric for image generation, utilizing deep learning-based embeddings to compare real and generated images in a feature space, capturing both fidelity and diversity \cite{heusel2017gans}. In the domain of audio, the Frechet Audio Distance (FAD) extended this approach to evaluate generative audio models by leveraging embeddings from pre-trained audio feature extractors \cite{kilgour2018fr}. 

Recent research has deepened the insight into the robustness of these Frechet distance based metrics, specifically highlighting the dependence on the choice of reference models and training datasets, which may introduce biases and limit generalizability \cite{gui2024adapting} \cite{tailleur2024correlation}. Nevertheless, an objective metric for generative evaluation should be able to reflect subtle, high-level attributes that humans intuitively recognize, such as style and expressiveness, while maintaining versatility and performance across various contexts.

\subsection{Evaluation of generative symbolic music}

When transitioning from images or audio to the evaluation of symbolic music generation, the task introduces new challenges. Music, as a structured and temporal art form, embodies intricate patterns in melody, harmony, and rhythm that are challenging to encapsulate with simple statistical or perceptual measures. Furthermore, the quality and characteristics of the audio signal itself have a significant role in the computation of FAD, leaving an uncertain portion of the evaluation to purely musical qualities, like style of composition. This complexity underscores the need for an evaluation metric that can effectively represent the multi-dimensional characteristics of music in a quantitative manner.

Existing objective metrics for evaluation of symbolic music often focus on low-level musical attributes, like the ratio of notes in scale \cite{dong2018musegan}, pitch class entropy \cite{wu2020jazz} or scale consistency \cite{mogren2016c}. Several such metrics are implemented in \texttt{MusPy} \cite{dong2020muspy}. While these metrics provide interesting and valuable insights into specific compositional aspects, they primarily capture surface-level characteristics. Furthermore, they do not always offer a comprehensive measure of higher-level musical qualities such as structure, coherence, and style, which are critical for assessing the overall musicality and perceived characteristics of generative outputs. Consequently, while a statistical comparison of these low-level metrics computed on a test set versus a reference set may give specific ideas of the musical traits of the test set, there has yet to be an agreement in how such a comparison helps to assess the overall quality of generative models.


\section{Frechet Music Distance}




\subsection{Symbolic music embedding models}

FMD leverages recent pioneering advancements in symbolic music representation learning, specifically the music encoders from CLaMP \cite{wu2023clamp} and CLaMP 2 \cite{wu2024clamp} which capture rich semantic musical representations of symbolic music. CLaMP is a contrastive music-text self-supervised model for
symbolic music representation learning. It was trained on 1.4 million music-text pairs
and tested on semantic search and classification tasks. It also introduced the
M3 music encoder for ABC symbolic music with bar patching, trained in a setting based on masked
autoencoders \cite{he2022masked}. CLaMP 2 includes MIDI support and
enhanced, multilingual text descriptions, allowing for improved alignment
between music and several languages. The music encoder in CLaMP 2 supports ABC and uses a interleaved representation for multi-track symbolic music. It also supports MIDI, utilizing an intermediate, lossless representation called MTF (MIDI Text
Format), which avoids common quantization issues often found in MIDI encoding.

\subsection{Data preprocessing}


For CLaMP 1, we use an ABC pre-processing pipeline adapted directly from \cite{wu2023clamp}.
In the case of MIDI and CLaMP 2, we convert MIDI to an intermediate representation of MTF (MIDI Text Format), which allows for lossless
representation of MIDI data as text.
In the case of ABC and CLaMP2, we find that representative and consistent results require specific preprocessing, especially for files from MIDI conversion and single-voice generative models. This involves removing empty spaces at the start of lines and adding voice information to enable voice-interleaved formatting, as outlined in \cite{wu2024clamp}. However, for single-track outputs from simple generative models, this processing offered minimal improvement and caused significant data loss, reducing dataset size. Consequently, we retained only the initial steps of cleaning empty spaces and adding voice information.

We use \texttt{abcmidi}, \texttt{MusPy} \cite{dong2020muspy} and \texttt{pretty\_midi} \cite{raffel2014intuitive} libraries for conversion between MIDI and ABC and for all other MIDI processing.

\begin{table*}[h]
\centering
\begin{tabular}{l l r r r}
\hline
\textbf{Reference Dataset} & \textbf{Generative Model} & \textbf{\boldmath$\mathrm{FMD}_{clamp2}^{midi}$} & \textbf{\boldmath$\mathrm{FMD}_{clamp2}^{abc}$} & \textbf{\boldmath$\mathrm{FMD}_{clamp1}^{abc}$} \\

\hline
Folk V2      & GPT-2       &  17.97        &      25.84     &  47.23         \\
Folk V2      & FolkRNN    &  105.09        &      91.24     &     59.42       \\ \hline
POP909       & GPT-2       &    751.41        & 522.44         & 199.63           \\
POP909       & FolkRNN    & 815.00             & 573.47      & 240.65                    \\
MAESTRO      & GPT-2       & 478.39    & 630.88         & 234.86          \\
MAESTRO      & FolkRNN    & 509.57        & 678.80       & 265.64             \\
MidiCaps     & GPT-2       & 449.90          & 499.17     & 180.64               \\
MidiCaps     & FolkRNN    & 565.15          & 573.63          & 214.46          \\

\hline
\end{tabular}
\caption{Overview of reference datasets and models trained only on the folk v2 dataset. We use all of the samples from the reference datasets to compute FMD.}
\label{tab:overview_reference_models}
\end{table*}

\section{Experimental Evaluation}

To evaluate FMD, we investigate several comparisons and compute three variants of FMD, with the subscript denoting the embedding model, and superscript denoting the modality: $\mathrm{FMD}_{clamp2}^{midi}$, $\mathrm{FMD}_{clamp2}^{abc}$, $\mathrm{FMD}_{clamp1}^{abc}$.

\subsection{Dataset benchmarks}

As a measure of initial validation of our approach, we compute FMD on subsets of \emph{the same} symbolic dataset. We use MAESTRO \cite{hawthorne2018enabling}, which is a dataset of virtuoso-level piano performances known for their high complexity and fluent rhythms, typical for classical piano music. It includes pieces by composers like Chopin, Liszt and Rachmaninoff. We also investigate MidiCaps \cite{melechovsky2024midicaps}, a recent large dataset of diverse MIDI paired with text descriptions. Table~\ref{tab:midicaps_subsets_fmd} presents FMD values for selected subsets of MidiCaps.

\begin{table}[h]
\centering
\begin{tabular}{l l r}
\hline
\textbf{Ref. set genres} & \textbf{Test set genres} & \textbf{\boldmath$\mathrm{FMD}_{clamp2}^{midi}$} \\
\hline
Amb., Elec. & Class., Pop & 104.47 \\
Amb., Elec. & Class., ST  & 97.53 \\
Amb., Elec. & Elec., Pop  & 33.05 \\
Amb., Elec. & Pop, Rock   & 56.94 \\
Amb., Elec. & Class., Elec. & 128.39 \\
Class., Pop & Class., ST  & 66.63 \\
Class., Pop & Elec., Pop  & 67.57 \\
Class., Pop & Pop, Rock   & 96.60 \\
Class., Pop & Class., Elec. & 102.27 \\
Class., ST  & Elec., Pop  & 94.83 \\
Class., ST  & Pop, Rock   & 136.20 \\
Class., ST  & Class., Elec. & 55.95 \\
Elec., Pop  & Pop, Rock   & 33.51 \\
Elec., Pop  & Class., Elec. & 100.23 \\
Pop, Rock   & Class., Elec. & 159.63 \\
\hline
\end{tabular}
\caption{FMD computed across various MidiCaps genre subsets. Each subset contains 10,000 samples. Abbreviations for genres: 
\textbf{Class.} = Classical, \textbf{ST} = Soundtrack, \textbf{Amb.} = Ambient, \textbf{Elec.} = Electronic.}
\label{tab:midicaps_subsets_fmd}
\end{table}

Table~\ref{tab:maestro-midicaps-table} shows $\mathrm{FMD}_{clamp2}^{midi}$ scores for subsets of MAESTRO and MidiCaps. Random sampling from MAESTRO yields consistently low FMD scores, reflecting its homogeneity in musical qualities, as it predominantly features music from a virtuoso-level piano competition with a classical repertoire.

Random sampling from MidiCaps results in consistently low FMD values. In contrast, testing MAESTRO against MidiCaps produces a high FMD of 378.28, indicating that the virtuoso piano style characteristic of MAESTRO is not commonly represented in MidiCaps. This observation is reinforced by testing a “classical piano” subset of MidiCaps against MAESTRO, which also produces a high FMD.


\begin{table*}[h]
\centering
\begin{tabular}{l l r r r}
\hline
\textbf{Reference set} & \textbf{Test set} & \textbf{Reference set size} & \textbf{Test size} & \textbf{\boldmath$\mathrm{FMD}_{clamp2}^{midi}$} \\
\hline
MAESTRO & MAESTRO Random Sample & 1276 & 1000 & 0.28 \\
MAESTRO & MAESTRO Random Sample & 1276 & 500 & 1.62 \\
MAESTRO & MAESTRO Random Sample & 1276 & 100 & 11.14 \\
MidiCaps & MidiCaps Random Sample & 168385 & 10000 & 1.15 \\
MidiCaps & MidiCaps Random Sample & 168385 & 1000 & 12.15 \\
MidiCaps & MAESTRO & 168385 & 1276 & 378.28 \\
MAESTRO & MidiCaps Classical Piano & 1276 & 4750 & 334.59 \\
\hline
\end{tabular}
\caption{FMD for MAESTRO and MidiCaps}
\label{tab:maestro-midicaps-table}
\end{table*}

\subsection{Evaluation of generative models}

We evaluate three symbolic generative models: MMT \cite{dong2023multitrack}, FolkRNN \cite{sturm2016music}, and a GPT-2 transformer model
\cite{radford2019language}. MMT is a recent multitrack MIDI transformer model. FolkRNN is a recurrent neural net originally used to generate folk music in the ABC format. The GPT-2 model is used for generating ABC music,
following a methodology similar to \cite{geerlings2020interacting}. The GPT-2 we
use for evaluation is a scaled down, minimal model with 4 blocks with 4
attention heads each and an embedding size of 256. It's trained on the same ABC
dataset as FolkRNN - we will call this dataset "folk v2" in following sections.

To ensure reproducibility, we utilize the implementations and pre-trained models of FolkRNN and GPT-2 provided by \texttt{Symbotunes} \cite{skiers2024symbotunes}. We sample a total of 10,000 tracks both from the GPT-2 and from the FolkRNN model. We compute $\mathrm{FMD}_{clamp2}^{midi}$, $\mathrm{FMD}_{clamp2}^{abc}$, and $\mathrm{FMD}_{clamp1}^{abc}$. In addition to evaluating the outputs of these models against their training set (folk v2), we also test them against external datasets \emph{not used} for training: MidiCaps, MAESTRO, and POP909 \cite{wang2020pop909}. Given the unique characteristics of the folk v2 dataset, we anticipate low FMD values for this training set and higher values for the other datasets, which is reflected in the results. These findings are presented in Table~\ref{tab:overview_reference_models}.

\begin{figure}[h]
        \centering
        \includegraphics[width=\linewidth]{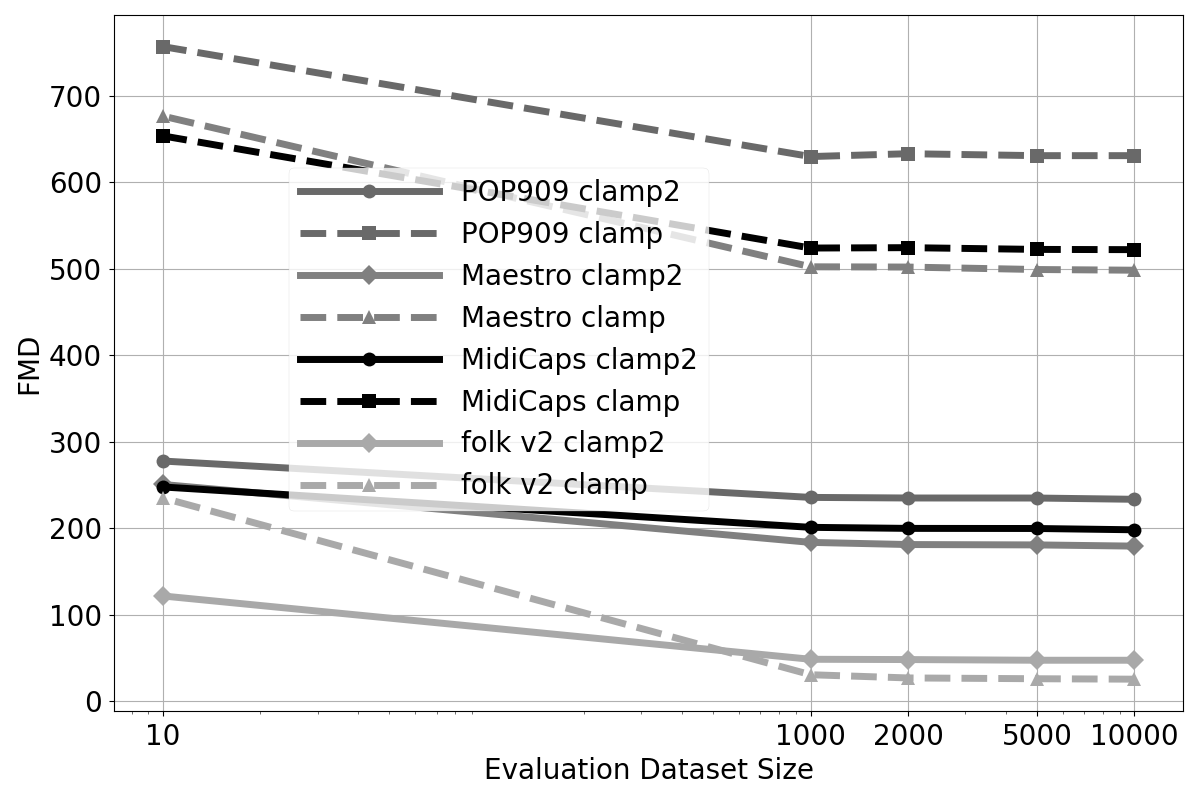}  %
        \caption{FMD for GPT-2, ABC evaluation.}
        \label{fig:abc_gpt2_baseline_compare}
    \end{figure}

\begin{figure}[h]
        \centering
        \includegraphics[width=\linewidth]{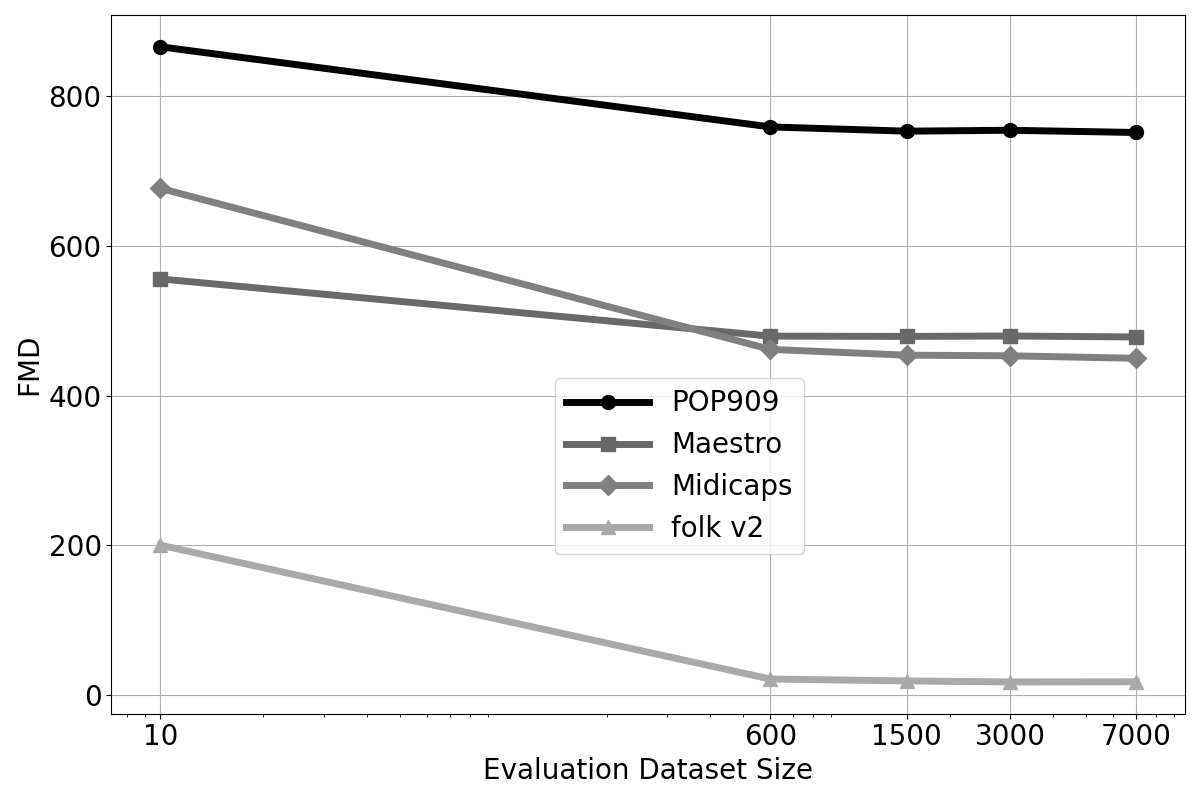}  %
        \caption{FMD for GPT-2, MIDI evaluation.}
        \label{fig:midi_gpt2_baseline_compare}
    \end{figure}

Figures~\ref{fig:abc_gpt2_baseline_compare} and \ref{fig:midi_gpt2_baseline_compare}, present a comparison of results depending on the evaluation sample size. We observe similar behavior and convergence of the FMD above a certain test set size for the FolkRNN model as well.

For the evaluation of MMT, we use the version trained on the Symphonic Orchestral Dataset (SOD) \cite{crestel2018database}. The original implementation of MMT provides four generation modes: \emph{unconditioned}, \emph{instrument informed}, where the model is given a sequence of instrument codes, and \emph{4-beat continuation} and \emph{16-beat continuation}, where the first $n$ beats are supplied to the model. We compute $\mathrm{FMD}_{clamp2}^{midi}$ for each of these modes on a set of 1000 generated samples, using 5710 samples from SOD as the reference set. The results, shown in Table~\ref{tab:mmt_sod_fmd}, reveal that $\mathrm{FMD}_{clamp2}^{midi}$ improves as additional constraints are applied to the model, aligning with the expectation that a model prompted with segments of the reference should produce outputs closer to it.

\begin{table}[h!]
\centering
\begin{tabular}{l l r}
\hline
 \textbf{Ref. set} & \textbf{MMT Generation Mode} & \textbf{\boldmath$\mathrm{FMD}_{clamp2}^{midi}$} \\
\hline
SOD & Unconditioned & 363.57 \\
SOD & Instrument Informed & 354.17 \\
SOD & 4-beat continuation & 335.77 \\
SOD & 16-beat continuation & 328.74 \\
\hline
\end{tabular}
\caption{FMD results for the MMT model, with the full SOD dataset serving as the the reference.}
\label{tab:mmt_sod_fmd}
\end{table}



\subsection{Outlier detection using FMD}

The unexpectedly high FMD values between the MAESTRO dataset, which features
classical piano music, and the MidiCaps songs tagged as classical and limited to
piano as the sole instrument, as reported in
Table~\ref{tab:maestro-midicaps-table}, prompted further investigation. To
explore this, we randomly sampled 10 songs from the MidiCaps subset, synthesized them,
and conducted a simple listening test. Although the test involved only five
participants, all were musicians. The participants were asked to decide \emph{whether the track they are listening to "is classical piano music"} and unanimously agreed that only
one of the sampled songs clearly aligned with that definition, suggesting
potential ambiguity in the classification of songs as “classical piano” within
the MidiCaps dataset.

\begin{figure}
    \centering
    \includegraphics[width=1\linewidth]{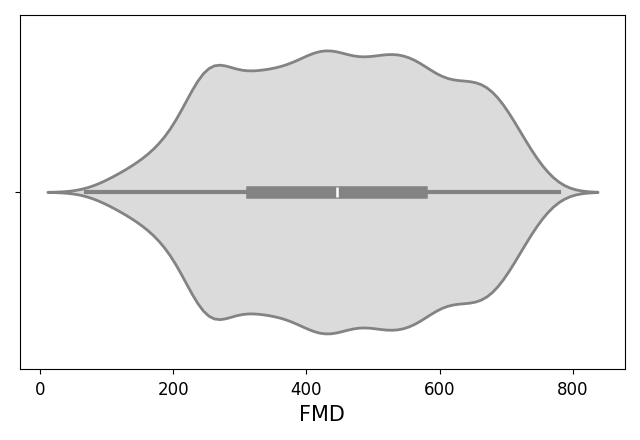}
    \caption{Distribution of per-song FMD between MAESTRO and Midicaps classical piano}
    \label{fig:maestro-midicaps-classical-piano}
\end{figure}

Subsequently, we computed the per-song FMD between MAESTRO and individual songs from the MidiCaps Classical Piano subset and created a new set containing only songs from the bottom 5th percentile of FMD. We then repeated the listening test with this new set. This time, the verdict was that all 10 samples were indeed classical piano music, suggesting that per-song FMD can effectively detect outliers differing from the reference set. The distribution of per-song FMD is shown on Figure~\ref{fig:maestro-midicaps-classical-piano}


\subsection{Musical sensitivity tests}

\begin{table*}[h]
\centering

\begin{tabular}{l l r r r}
\hline
\textbf{Reference Dataset} & \textbf{Model} & \textbf{\boldmath$\mathrm{FMD}_{clamp2}^{midi}$} & \textbf{\boldmath$\mathrm{FMD}_{clamp2}^{abc}$} & \textbf{\boldmath$\mathrm{FMD}_{clamp1}^{abc}$} \\

\hline
folk v2 CMaj  & GPT-2 &  33.066  &  27.40 &  51.92 \\
folk v2  CMaj & FolkRNN    &  83.35  & 81.95  &  58.00 \\

\hline
\end{tabular}
\caption{Musical mode impact on FMD values. All FMD scores for FolkRNN have improved when computed against a reference dataset containing only C Major tunes. \textbf{\boldmath$\mathrm{FMD}_{clamp2}^{midi}$} for GPT2 worsens, while two other FMD measures stay within a difference of $2.5$.}
\label{tab:cmaj_only}
\end{table*}

\begin{table}[h]
\centering
\begin{tabular}{l l l r r r}
\hline
\textbf{Baseline} & \textbf{Augmented} & \textbf{p} & \textbf{\boldmath$\mu$} & \textbf{\boldmath$\sigma$} & \textbf{\boldmath$\mathrm{FMD}_{clamp2}^{midi}$} \\
\hline
MAESTRO & Note Pitch & 0.01 & 0 & 5 & 0.07 \\
MAESTRO & Note Pitch & 0.10 & 0 & 5 & 4.86 \\
MAESTRO & Note Pitch & 0.25 & 0 & 5 & 26.29 \\
MAESTRO & Note Pitch & 0.50 & 0 & 5 & 82.52 \\
MAESTRO & Note Pitch & 0.75 & 0 & 5 & 130.40 \\
MAESTRO & Note Pitch & 0.90 & 0 & 5 & 146.35 \\
MAESTRO & Note Pitch & 0.01 & 0 & 10 & 0.25 \\
MAESTRO & Note Pitch & 0.10 & 0 & 10 & 13.93 \\
MAESTRO & Note Pitch & 0.25 & 0 & 10 & 56.82 \\
MAESTRO & Note Pitch & 0.50 & 0 & 10 & 136.66 \\
MAESTRO & Note Pitch & 0.75 & 0 & 10 & 193.87 \\
MAESTRO & Note Pitch & 0.90 & 0 & 10 & 213.31 \\
\hline
\end{tabular}
\caption{FMD results for the MAESTRO dataset with additive noise. The dataset consists of 1,276 entries. The parameter \textbf{p} denotes the probability of a note being modified. The parameters \textbf{\boldmath$\mu$} and \textbf{\boldmath$\sigma$} define the normal distribution from which the noise added to the note is sampled.}
\label{tab:augmentations}
\end{table}


The original folk v2 dataset \cite{sturm2016music} comprises transcriptions in four musical modes: major, minor, dorian, and mixolydian, with the major mode being the most prevalent, appearing in 67\% of the samples. Analysis reveals that our FolkRNN model exclusively generates samples in the dominant mode of C Major, whereas the GPT-2 model produces outputs across all modes. To examine whether FMD captures this characteristic, we use a reference subset consisting solely of C Major samples from folk v2. Compared to the results in Table~\ref{tab:overview_reference_models}, we expect improved results for FolkRNN and potentially diminished results for GPT-2. The outcomes of this experiment are presented in Table~\ref{tab:cmaj_only}.

Another experiment involved augmenting MIDI tracks with random noise. For each note, there was a probability $p$ of it being augmented. The augmentation consisted of adding Gaussian noise to a property of the MIDI note. We tested augmentations of both Note Pitch and Note Velocity. Since pitch and velocity are discrete values within the range [0, 127], the noise was rounded to the nearest integer and bounded to ensure it remained within this valid range. These augmented datasets were compared against the original, non-augmented reference dataset. The FMD for velocity augmentations was approximately 0, indicating that FMD is insensitive to this value. For pitch augmentations, the results are shown in Table~\ref{tab:augmentations}. They seem to have a significant impact on value of FMD.

\subsection{Feature distribution estimation methods}

To make the metric more robust when dealing with small sample sizes, we evaluated several mean and covariance estimation methods. The methods tested include Bootstrapping, Basic Shrinkage, Ledoit-Wolf Shrinkage \cite{ledoit2004well}, and Oracle Approximating Shrinkage (OAS) \cite{Chen_2010}. We also attempted to utilize Graphical Lasso \cite{friedman2008sparse} and Minimum Covariance Determinant \cite{rousseeuw1984least} \cite{rousseeuw1999fast}; however, both methods encountered numerical stability issues, likely due to covariance matrices being near singular. The MLE was implemented in the same manner as in the FAD, using \texttt{NumPy} \cite{harris2020array}. The Bootstrap estimator implemented by us. For all other estimators we used \texttt{Scikit-learn} \cite{scikit-learn}.

\begin{figure}
    \centering
    \includegraphics[width=1\linewidth]{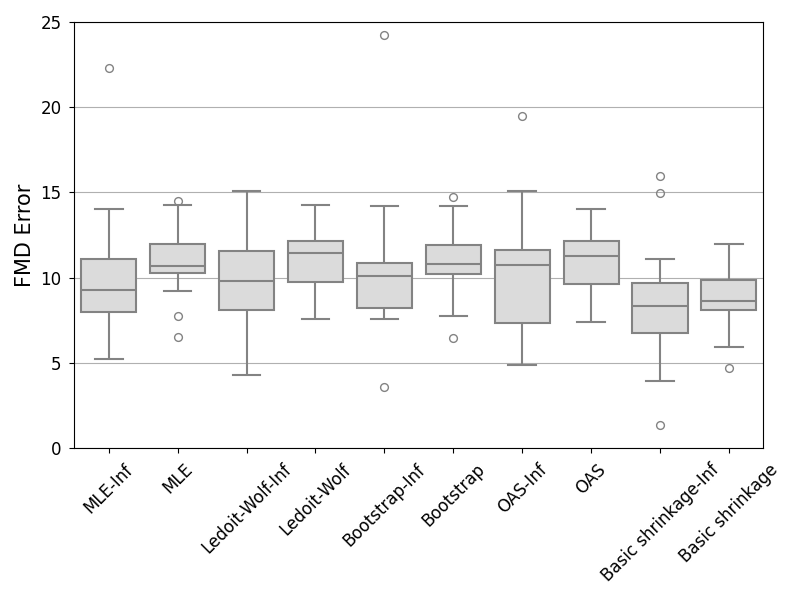}
    \caption{FMD errors for different mean and covariance estimation methods computed on combinations of MidiCaps subsets.}
    \label{fig:estimation_methods_1000_all_genres}
\end{figure}

The selected methods were benchmarked against the baseline Maximum Likelihood Estimator (MLE). CLaMP 2 was used as the embedding model. These methods were evaluated in the context of both the regular FMD and FMD-inf metrics. For the evaluations, we utilized subsets of the MidiCaps dataset, categorized according to tagged genres. Each genre class had two subsets: one with 10,000 samples and another with 1,000 samples, both obtained through sampling without replacement. For each pair of genre classes, FMD was calculated using a reference subset of size 10,000 from one class and a test subset of size 10,000 from another class. This value served as the ground truth. Subsequently, FMD was calculated between the same reference subset and a smaller version of the previous test subset within the same genre category, containing 1,000 samples. The objective was to minimize the discrepancy between FMD values when using datasets of different sizes.
The results can be seen on Figure~\ref{fig:estimation_methods_1000_all_genres}

\begin{figure}
    \centering
    \includegraphics[width=1\linewidth]{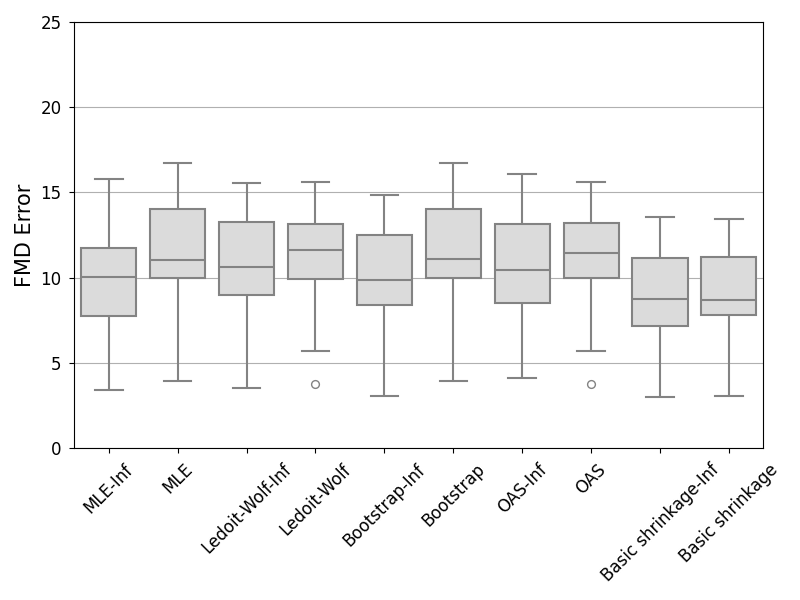}
    \caption{FMD errors for different mean and covariance estimation methods computed only for midicaps samples that came from the same subset.}
    \label{fig:estimation_methods_same_genres}
\end{figure}

We also conducted a second experiment, computing FMD only between subsets from the same genre class. This approach allowed us to know the precise ground truth FMD value of zero, eliminating bias from estimating the ground truth using subsets of size 10,000. To make the experiment more representative, we included another group of subsets of size 2,000, compensating for the reduced number of combinations compared to the previous experiment. The results are shown on Figure~\ref{fig:estimation_methods_same_genres}

Both experiments yielded similar results. While the estimation methods were comparable and all tended to overestimate the FMD for small sample sizes, the Basic Shrinkage method emerged as the best performer. This method, both in its standard form and with infinite sample size extrapolation, consistently produced the smallest FMD values and appeared to be the least affected by smaller sample sizes. However, it is worth noting that the $R^2$ value for infinite sample size extrapolation was highest for MLE, suggesting a better regression fit.

\section{Limitations and further work}

While FMD provides a new metric for assessing the similarity between generated and reference music distributions, we take note of its limitations that warrant further investigation. First, FMD’s reliance on embedding models such as CLaMP2 introduces a dependency on the pretraining data and architecture of these models, which may bias the metric toward certain musical styles or genres. Additionally, the sensitivity of FMD to subtle musical variations, such as mode shifts or instrument choices, is not fully understood, as highlighted by experiments with subsets of datasets like folk v2 and MidiCaps. The metric may also struggle with ambiguous genre classifications or datasets with heterogeneous tagging systems, potentially leading to misleadingly high or low scores - however, by doing so, it may also highlight inconsistencies and provide insights that could drive improvements in dataset quality and tagging practices. 

Several experiments highlight the critical importance of data pre-processing, particularly for ABC data. Even minor modifications in data formatting can have a substantial effect on the outcomes. Each pre-processing step is also sensitive to the integrity of the information within the file. Fully utilizing a voice interleaved form \cite{wu2024clamp} could potentially enhance some results, but we have found it frequently encountered errors when the input data was in any way corrupted or lacked the necessary information for interleaving. 

Future work could focus on investigating and refining the FMD to account for temporal and structural elements of music. This includes developing task-specific variants and deepening the analysis of musical content within FMD. Furthermore, comparing FMD with FAD, specifically examining the correlations and differences across the same datasets in different modalities may deepen the insight into particular musical traits and characteristics. 

Lastly, extensive subjective validation through listening tests with musicians and other individuals possessing musical expertise should complement FMD to ensure that its quantitative assessments align with human perceptions of musical similarity.

\section{Conclusion}

We introduce the Frechet Music Distance (FMD), a specialized adaptation of the Frechet family of metrics designed to evaluate generative symbolic music while accounting for abstract musical characteristics. We evaluate it on ABC and MIDI representations with current state of the art symbolic music embedding models. We also analyze and highlight some of the limitations of FMD. 

By introducing FMD, we aim to establish a reproducible standard for the quantitative evaluation of generative models in symbolic music, facilitating more nuanced comparisons and supporting advancements in the field. 


\bibliography{aaai25}

\end{document}